\documentstyle[12pt]{article}
\advance \topmargin by -\headheight
\advance \topmargin by -\headsep
     
\evensidemargin \oddsidemargin
\begin{document}

\def\rf#1{(\ref{eq:#1})}
\def\lab#1{\label{eq:#1}}
\def\nonu{\nonumber}
\def\eq{\!\!\!\! &=& \!\!\!\! }
\def\br{\begin{eqnarray}}
\def\er{\end{eqnarray}}
\def\be{\begin{equation}}
\def\ee{\end{equation}}
\def\lb{\lbrack}
\def\rb{\rbrack}
\def\llangle{\left\langle}
\def\rrangle{\right\rangle}
\def\blangle{\Bigl\langle}
\def\brangle{\Bigr\rangle}
\def\llb{\left\lbrack}
\def\rrb{\right\rbrack}
\def\lcurl{\left\{}
\def\rcurl{\right\}}
\def\({\left(}
\def\){\right)}
\def\v{\vert}                     
\def\bv{\bigm\vert}               
\def\lskip{\vskip\baselineskip\vskip-\parskip\noindent}
\def\mskp{\par\vskip 0.3cm \par\noindent}
\def\tr{\mathop{\rm tr}}                  
\def\Tr{\mathop{\rm Tr}}                  
\newcommand\partder[2]{{{\partial {#1}}\over{\partial {#2}}}}
\newcommand\Bil[2]{\Bigl\langle {#1} \Bigg\vert {#2} \Bigr\rangle}  
\newcommand\bil[2]{\left\langle {#1} \bigg\vert {#2} \right\rangle} 
\newcommand\me[2]{\left\langle {#1}\right|\left. {#2} \right\rangle} 
\newcommand\sbr[2]{\left\lbrack\,{#1}\, ,\,{#2}\,\right\rbrack} 
\newcommand\Sbr[2]{\Bigl\lbrack\,{#1}\, ,\,{#2}\,\Bigr\rbrack}
\newcommand\pbr[2]{\{\,{#1}\, ,\,{#2}\,\}}       
\newcommand\Pbr[2]{\Bigl\{ \,{#1}\, ,\,{#2}\,\Bigr\}}
\newcommand\pbbr[2]{\lcurl\,{#1}\, ,\,{#2}\,\rcurl}
\newcommand\ttmat[9]{\left(\begin{array}{ccc}  
{#1} & {#2} & {#3} \\ {#4} & {#5} & {#6} \\
{#7} & {#8} & {#9} \end{array} \right)}
\newcommand\thrcol[3]{\left(\begin{array}{c}  
{#1} \\ {#2} \\ {#3} \end{array} \right)}
\def\bp{{\bar \p}}
\def\tit{{\tilde t}}
\def\a{\alpha}
\def\b{\beta}
\def\c{\chi}
\def\d{\delta}
\def\D{\Delta}
\def\eps{\epsilon}
\def\vareps{\varepsilon}
\def\g{\gamma}
\def\G{\Gamma}
\def\grad{\nabla}
\def\h{{1\over 2}}
\def\l{\lambda}
\def\L{\Lambda}
\def\m{\mu}
\def\n{\nu}
\def\o{\over}
\def\om{\omega}
\def\O{\Omega}
\def\p{\phi}
\def\P{\Phi}
\def\pa{\partial}
\def\pr{\prime}
\def\ra{\rightarrow}
\def\lra{\longrightarrow}
\def\s{\sigma}
\def\S{\Sigma}
\def\t{\tau}
\def\th{\theta}
\def\Th{\Theta}
\def\z{\zeta}
\def\ti{\tilde}
\def\wti{\widetilde}
\newcommand\sumi[1]{\sum_{#1}^{\infty}}
\newcommand\fourmat[4]{\left(\begin{array}{cc}  
{#1} & {#2} \\ {#3} & {#4} \end{array} \right)}
\newcommand\twocol[2]{\left(\begin{array}{cc}  
{#1} \\ {#2} \end{array} \right)}
\def\cKP{{\sf cKP}~}
\def\bfs{{\bf s}}
\def\bfts{{\bf {\tilde s}}}
\def\qs{Q_{\bfs}}
\def\kere{\mbox{\rm Ker (ad $E$)}}
\def\ime{\mbox{\rm Im (ad $E$)}}
\def\cgh{{\widehat {\cal G}}}
\def\vp{\varphi}
\def\qsp{Q_{\bfs^{\pr}}}
\def\bfsp{{\bf s}^{\prime}}
\newcommand\ket[1]{\vert {#1}\rangle}

\font\numbers=cmss12
\font\upright=cmu10 scaled\magstep1
\def\stroke{\vrule height8pt width0.4pt depth-0.1pt}
\def\topfleck{\vrule height8pt width0.5pt depth-5.9pt}
\def\botfleck{\vrule height2pt width0.5pt depth0.1pt}
\def\Zmath{\vcenter{\hbox{\numbers\rlap{\rlap{Z}\kern 0.8pt\topfleck}\kern
2.2pt
                   \rlap Z\kern 6pt\botfleck\kern 1pt}}}
\def\Qmath{\vcenter{\hbox{\upright\rlap{\rlap{Q}\kern
                   3.8pt\stroke}\phantom{Q}}}}
\def\Nmath{\vcenter{\hbox{\upright\rlap{I}\kern 1.7pt N}}}
\def\Cmath{\vcenter{\hbox{\upright\rlap{\rlap{C}\kern
                   3.8pt\stroke}\phantom{C}}}}
\def\Rmath{\vcenter{\hbox{\upright\rlap{I}\kern 1.7pt R}}}
\def\IZ{\ifmmode\Zmath\else$\Zmath$\fi}
\def\IQ{\ifmmode\Qmath\else$\Qmath$\fi}
\def\IN{\ifmmode\Nmath\else$\Nmath$\fi}
\def\IC{\ifmmode\Cmath\else$\Cmath$\fi}
\def\IR{\ifmmode\Rmath\else$\Rmath$\fi}


\def\cA{{\cal A}}
\def\cB{{\cal B}}
\def\cC{{\cal C}}
\def\cD{{\cal D}}
\def\cE{{\cal E}}
\def\cF{{\cal F}}
\def\cG{{\cal G}}
\def\cH{{\cal H}}
\def\cI{{\cal I}}
\def\cJ{{\cal J}}
\def\cK{{\cal K}}
\def\cL{{\cal L}}
\def\cM{{\cal M}}
\def\cN{{\cal N}}
\def\cO{{\cal O}}
\def\cP{{\cal P}}
\def\cQ{{\cal Q}}
\def\cR{{\cal R}}
\def\cS{{\cal S}}
\def\cT{{\cal T}}
\def\cU{{\cal U}}
\def\cV{{\cal V}}
\def\cW{{\cal W}}
\def\cY{{\cal Y}}
\def\cZ{{\cal Z}}
\def\one{\hbox{{1}\kern-.25em\hbox{l}}}
\def\0#1{\relax\ifmmode\mathaccent"7017{#1}%
        \else\accent23#1\relax\fi}
\newtheorem{definition}{Definition}[section]
\newtheorem{proposition}{Proposition}[section]
\newtheorem{theorem}{Theorem}[section]
\newtheorem{lemma}{Lemma}[section]
\newtheorem{corollary}{Corollary}[section]
\def\proof{\par{\it Proof}. \ignorespaces} \def\endproof{{$\Box$}\par}
\newenvironment{Proof}{\proof}{\endproof}
\def\cKP{{\sf cKP}~}
\newcommand\DB{{Darboux-B\"{a}cklund}~}
\newcommand{\nit}{\noindent}
\newcommand{\ct}[1]{(\cite{#1})}
\newcommand{\bi}[1]{\bibitem{#1}}
%
%
\newcommand\PRL[3]{{\sl Phys. Rev. Lett.} {\bf#1} (#2) #3}
\newcommand\NPB[3]{{\sl Nucl. Phys.} {\bf B#1} (#2) #3}
\newcommand\NPBFS[4]{{\sl Nucl. Phys.} {\bf B#2} [FS#1] (#3) #4}
\newcommand\CMP[3]{{\sl Commun. Math. Phys.} {\bf #1} (#2) #3}
\newcommand\PRD[3]{{\sl Phys. Rev.} {\bf D#1} (#2) #3}
\newcommand\PLA[3]{{\sl Phys. Lett.} {\bf #1A} (#2) #3}
\newcommand\PLB[3]{{\sl Phys. Lett.} {\bf #1B} (#2) #3}
\newcommand\JMP[3]{{\sl J. Math. Phys.} {\bf #1} (#2) #3}
\newcommand\PTP[3]{{\sl Prog. Theor. Phys.} {\bf #1} (#2) #3}
\newcommand\SPTP[3]{{\sl Suppl. Prog. Theor. Phys.} {\bf #1} (#2) #3}
\newcommand\AoP[3]{{\sl Ann. of Phys.} {\bf #1} (#2) #3}
\newcommand\RMP[3]{{\sl Rev. Mod. Phys.} {\bf #1} (#2) #3}
\newcommand\PR[3]{{\sl Phys. Reports} {\bf #1} (#2) #3}
\newcommand\FAP[3]{{\sl Funkt. Anal. Prilozheniya} {\bf #1} (#2) #3}
\newcommand\FAaIA[3]{{\sl Functional Analysis and Its Application} {\bf #1}
(#2) #3}
\def\TAMS#1#2#3{{\sl Trans. Am. Math. Soc.} {\bf #1} (#2) #3}
\def\InvM#1#2#3{{\sl Invent. Math.} {\bf #1} (#2) #3}
\def\AdM#1#2#3{{\sl Advances in Math.} {\bf #1} (#2) #3}
\def\PNAS#1#2#3{{\sl Proc. Natl. Acad. Sci. USA} {\bf #1} (#2) #3}
\newcommand\LMP[3]{{\sl Letters in Math. Phys.} {\bf #1} (#2) #3}
\newcommand\IJMPA[3]{{\sl Int. J. Mod. Phys.} {\bf A#1} (#2) #3}
\newcommand\TMP[3]{{\sl Theor. Mat. Phys.} {\bf #1} (#2) #3}
\newcommand\JPA[3]{{\sl J. Physics} {\bf A#1} (#2) #3}
\newcommand\JSM[3]{{\sl J. Soviet Math.} {\bf #1} (#2) #3}
\newcommand\MPLA[3]{{\sl Mod. Phys. Lett.} {\bf A#1} (#2) #3}
\newcommand\JETP[3]{{\sl Sov. Phys. JETP} {\bf #1} (#2) #3}
\newcommand\JETPL[3]{{\sl  Sov. Phys. JETP Lett.} {\bf #1} (#2) #3}
\newcommand\PHSA[3]{{\sl Physica} {\bf A#1} (#2) #3}
\newcommand\PHSD[3]{{\sl Physica} {\bf D#1} (#2) #3}
\newcommand\JPSJ[3]{{\sl J. Phys. Soc. Jpn.} {\bf #1} (#2) #3}
\newcommand\JGP[3]{{\sl J. Geom. Phys.} {\bf #1} (#2) #3}

\begin{titlepage}
\vspace*{-2 cm}
\noindent
August, 1997 \hfill{IFT-P.053/97}\\
solv-int/9709004 \hfill{UICHEP-TH/97-7}

\vskip 3cm

\begin{center}
{\Large\bf Solitons from Dressing in an Algebraic Approach \\
to the Constrained KP Hierarchy}
\vglue 1  true cm
H. Aratyn$^1$ , L.A. Ferreira$^2$, J.F. Gomes$^2$ and A.H. Zimerman$^2$\\

\vspace{1 cm}
$^1${\footnotesize Department of Physics \\
University of Illinois at Chicago\\
845 W. Taylor St., Chicago, IL 60607-7059}

\vspace{1 cm}

$^2${\footnotesize Instituto de F\'\i sica Te\'orica - IFT/UNESP\\
Rua Pamplona 145\\
01405-900, S\~ao Paulo - SP, Brazil}\\

\medskip
\end{center}

\normalsize
\vskip 0.2cm

\begin{abstract}

The algebraic matrix hierarchy approach based on affine Lie $sl (n)$
algebras leads to a variety of $1+1$ soliton equations.
By varying the rank of the underlying $sl (n)$ algebra as well as its
gradation in the affine setting, one encompasses the set of the soliton
equations of the constrained KP hierarchy.

The soliton solutions are then obtained as elements of the orbits of the
dressing transformations constructed in terms of representations of the
vertex operators of the affine $sl (n)$ algebras realized in
the unconventional gradations.
Such soliton solutions exhibit non-trivial dependence on
the KdV (odd) time flows and KP (odd and even) time flows
which distinguishes them from the conventional structure of the
Darboux-B\"{a}cklund Wronskian solutions of the constrained KP hierarchy.
\end{abstract}
\end{titlepage}

\section{Introduction. The Algebraic cKP Model}
A large class of $1+1$ soliton equations belongs to the so-called constrained
KP (cKP) hierarchy.
Some of the most prominent members of this group are the KdV and
the nonlinear Schr\"{o}dinger equations of the AKNS model.
The cKP evolution equations possess the familiar Lax pair representations
with generally pseudo-differential Lax operators which emerge naturally
as reductions of the complete KP hierarchy Lax operators \ct{ckp}.
Conventionally, the constrained KP hierarchy is obtained from the KP
hierarchy by a process of reduction involving the so-called eigenfunctions.
The eigenfunctions appear in the constraint relations introducing a
functional dependence between initially infinitely many coefficients of
the KP Lax operator.
This scheme results in the pseudo-differential cKP Lax operator
of the type $\cL = L_{K+1} + \sum_{i=1}^M \P_i \pa^{-1} \Psi_i$, where
$L_{K+1}$ is the differential operator of the $K+1$-order, while
$\P_i, \Psi_i$ are the eigenfunctions of $\cL$.
In general, $\cL$ possesses a finite number of coefficients which enter the
soliton equations and depend on all $(t_1,t_2, t_3, \ldots )$
isospectral time flows of the KP hierarchy.
In the special $M=0$ case with the cKP Lax operator being a purely differential
operator  $\cL = L_{K+1}$ we encounter dependence on only some
of the original time-flows of the KP hierarchy. The most simple example
($K=1, M=0$) is the KdV hierarchy with only odd time flows present.

The soliton solutions for the cKP models have been found
in \ct{ANP2} for the arbitrary $K$ and $M=1$ case using the
Darboux-B\"{a}cklund technique.
Generalization to an arbitrary $M$ is simple and was given in
\ct{ANP3} (see also \ct{new}).
These solutions appeared in the Wronskian form in terms
of the eigenfunctions of the ``undressed'' $\cL = \pa^{K+1}$ Lax operator.

Here, we will present an alternative algebraic viewpoint of the
constrained KP hierarchy.
In this setting the algebraic dressing methods will provide new
soliton solutions which appear to differ from the conventional
form of the Darboux-Backlund Wronskian solutions due to a non-trivial
mixing of the KdV-like versus KP time flows. This will be shown explicitly
in the example characterized by $K=M=1$.

In an algebraic approach to the constrained KP hierarchy \ct{AFGZ}
the soliton evolution equations emerge as integrability conditions
of the following matrix eigenvalue problem:
\be
L \Psi = 0 \quad ;\quad L \equiv (D - A  -  E ) \quad; \quad
D \equiv I \partder{}{x}
\lab{lpsi}
\ee
with the matrix Lax operator $L$ belonging to Kac-Moody algebra
${\hat \cG}= {\widehat {sl}} (M+ K+1)$.
The integrable hierarchy is determined by the choice of gradation of
${\hat \cG}$.
By varying the Kac-Moody algebras together with their gradations one is able
to reproduce from the matrix hierarchy of eq. \rf{lpsi} the nonlinear
evolution equations of the cKP hierarchy.

We will be working with a simple setting in which the
matrix $E$ in eq.\rf{lpsi} has gradation $1$ with respect to gradation
specified by the vector \ct{kac}:
\be
\bfs = ( 1, \underbrace{0, \ldots ,0}_{M}, \underbrace{1, \ldots,1}_{K}\, )
\lab{svector}
\ee
We call this gradation an intermediate gradation as it interpolates
between the principal $ \bfs_{\rm principal} = (1, 1,\ldots , 1)$
and the homogeneous one $ \bfs_{\rm homogeneous} = (1,0 \ldots , 0)$.
As it is well-known the Wilson-Drinfeld-Sokolov
\ct{D-S,wilson,GIH1,GIH2,mcintosh} procedure
gives, respectively, the (m-)KdV \ct{GIH1} and AKNS \ct{FK83,AGZ} hierarchies
in these two limits.

Alternatively, the gradation $\bfs$ can be specified by an operator:
\be
\qs \equiv \sum_{a=1}^{K}  \l_{M+a} \cdot H^0 + (K+1) d
\lab{grading}
\ee
Here $\l_j$ are the fundamental weights and $d$ is the standard loop algebra
derivation.
Correspondingly, $E$ stands for
\be
E = \sum_{a=1}^{K}  E^{(0)}_{\a_{M+a}}
+   E^{(1)}_{-(\a_{M+1}+ \cdots+\a_{M+K})}
\lab{ele}
\ee
It is a non-regular (for $M >0$) and semisimple element of ${\hat \cG}$.

The matrix $A$ in eq. \rf{lpsi} contains the dynamical variables of the
model. $A$ is such that it has gradation zero and is parametrized in terms
of the dynamical variables $q_i$, $r_i$, $U_a$ and $\nu$ as follows:
\be
A  = \sum_{i=1}^{M} \( q_i P_i + r_i {P}_{-i} \) +
\sum_{a=1}^{K} U_{M+a} ( \a_{M+a} \cdot H^{(0)}) + \nu \, {\hat c}
\lab{a20}
\ee
where
$P_{\pm i} = E_{\pm (\a_{i} + \a_{i+1} + \ldots +\a_{M})}^{(0)}\, ,
\,\, i=1, 2, \ldots, M$,
and ${\hat c}$ is a central element of ${\hat \cG}$.

\section{The Heisenberg Subalgebra and the Vertex Operator}

For a regular element $E$ in the conventional Drinfeld-Sokolov
approach the isospectral flows are associated with the Heisenberg
subalgebra which can be identified with $\kere$.
Here, due to non-regularity of $E$  the Heisenberg algebra is
associated to the center of $\kere$.
It consists of the following three separate sets of operators:

\nit
1)~ \underbar{A homogeneous part of ${\hat {sl}} (M)$}
$\quad, \quad i=1,2, \ldots, M-1$
\be
\cK^{(n)}_i \, = \, {\sum_{p=1}^i p \, \a_p \cdot H^{(n)}  \o N_i}
\qquad; \qquad  N_i \equiv \sqrt{i (i+1)}
\lab{knl}
\ee

\nit
2)~ \underbar{A principal part of ${\hat {sl}} (K+1)$}
$\quad , \quad a=1,2, \ldots  ,K$
\be
{\cal A}^a_{a+n(K+1)} = \sum _{i=1}^{K+1-a}
E^{(n)}_{\a _{i+M}+ \a _{i+M+1} + \cdots +\a _{i+M+a-1}}
 + \sum _{i=1}^{a} E^{(n+1)}_{-(\a _{i+M}+ \a _{i+M+1}
+ \cdots +\a _{i+M+K-a})}
\lab{alpha}
\ee
\nit
3)~ \underbar{``A border term''}
\be
{\cal A}^0_{n(K+1)} = \sqrt{M+K+1 \o M} \, \l_M . H^{(n)}  - {K \o 2}
\, \sqrt{M \o M+K+1} \,{\hat c}\, \d_{n,0}
\lab{alphaz}
\ee

These relations provide a parametrization of the Heisenberg subalgebra in
terms of elements:
\be
b_{N,a} \equiv {\cal A}^a_{N=a+n(K+1)}\;\, , \;\, b_{N,0}
\equiv {\cal A}^0_{N=n(K+1)}\;\, ,\; \,
b_{N,i} \equiv {\cal K}_i^{(N)}
\lab{heisel}
\ee
where $a=1,2, \ldots, K$, $ i=1,2,\ldots ,M-1$.
The Heisenberg subalgebra elements from \rf{heisel} enter the oscillator
algebra relations (we put $c=1$):
\br
\sbr{b_{N,a}}{b_{N^{\pr},b}} &=& N \d_{N+N^{\pr} } \d_{a, K+1-b} \quad ;\;\;
a,b=1,2, \ldots, K \lab{osca}\\
\sbr{b_{N,0}}{b_{N^{\pr},0}} &=& N \d_{N+N^{\pr} } \lab{osca1}\\
\sbr{b_{N,i}}{b_{N^{\pr},j}} &=& N \d_{N+N^{\pr} }\, \d_{ij}\quad ;\;\;
i,j = 1,2,\ldots , M-1 \lab{osca2}
\er
Define next the Fubini-Veneziano operators:
\br
Q_{1 \leq i \leq M-1} (z) &=& 
i \sumi{n=1} { \cK^{(n)}_i z^{-n} \o n}
\quad ;\quad Q_M (z) = 
i \sumi{n=1} { {\cal A}^0_{n(K+1)} z^{-n(K+1)}
\o n(K+1)}  \lab{qi}\\ 
Q_{M+a} (z) &=& i \sumi{n=0} { {\cal A}^a_{a+n(K+1)} z^{a+n(K+1)}
  \o a+ n(K+1)}
\quad ;\quad a=1,2,\ldots, K \lab{qma}
\er
The corresponding conjugated Fubini-Veneziano operators $Q^{\dag} (z)$ are
obtained from \rf{qi}-\rf{qma} by taking into consideration rules
${ \cK^{(n)}}^{\dag}_i = \cK^{(-n)}_i$,
$\({\cal A}^0_{n(K+1)}\)^{\dag}={\cal A}^0_{-n(K+1)}$,
$\({\cal A}^a_{a+n(K+1)}\)^{\dag} = {\cal A}^a_{a-n(K+1)}$
as well as $z^{\dag} = z^{-1}$.


In reference \ct{workshop} we found the step operators of $sl (M+K+1)$
associated with the Cartan subalgebra defined by the Heisenberg subalgebra
\rf{heisel}.
That in turn enabled us to find the corresponding simple root structure for
$sl (M+K+1)$ with the intermediate grading.

Knowledge of roots and the Fubini-Veneziano operators is all what is needed
to write down a compact expression for the general vertex operator
in the normal ordered form:
\be
V^{\a} (z)\; \equiv \;z^{\h \sum_{j=1}^M (\a^j)^2}\,
\exp \({i {\vec \a}^{\ast} \cdot {\vec Q}^{\dag} (z)}\)\,
\exp \({i {\vec \a} \cdot {\vec q}}\)\, \exp \({{\vec \a} \cdot {\vec p} \ln
z}\) \, \exp \({i {\vec \a}\cdot {\vec Q} (z)}\)
\lab{vertex}
\ee
with $M+K$-component root vector ${\vec \a}$ described in \ct{workshop}
and $M+K$-component  vector ${\vec Q}$ having components described in
\rf{qi}-\rf{qma}. The zero-mode vectors ${\vec p}$ and ${\vec q}$
have only first $M$ components different from zero according to
$({\vec p})_i = p_i \, \th (M-i)$ and $({\vec q})_i = q_i \, \th (M-i)$.
They satisfy relations $\sbr{p_i}{q_j}= - i \d_{ij}$.
Furthermore, $p_M$ is equal to ${\cal A}^0_{n=0}$ from expression \rf{alphaz}.

An explicit example of the $sl(3)$ vertex will be given below in section 4.

\section{The Dressing Technique and the Tau-function}
The dressing technique \ct{FMSG} deals with reproducing of the nontrivial
part $E+A$ of the Lax matrix operator from eq. \rf{lpsi} by the gauge
transformations involving generators of positive and negative gradings
applied to the semisimple element $E$:
\br
E+A &= & \Th \,E\, \Th^{-1} + \(\pa_x \Th\) \Th^{-1}     \lab{tha}\\
E+A &= & \(B^{-1} \G\)  \,E\, \(\G^{-1} B\) + \(\pa_x B^{-1} \G\)
\(\G^{-1} B\) \lab{bgamma}
\er
where $B^{-1} \G$ contains positive terms and $\Th$ is an expansion in the
terms of negative grading such that
$ \Th = \exp \( \sum_{l <0} \th^{(l)} \) = 1 + \th^{(-1)} + \ldots $.
{}From expressions \rf{tha} and \rf{bgamma} we obtain two alternative
formulas for the same term $A$ of grade $0$:
\be
A =  - \sbr{E}{\th^{(-1)}} \; \;\quad {\rm or} \; \;\quad
A =  - B^{-1} \(\pa_x B \) \lab{abb}
\ee
The term $\th^{(-1)}$ of grade $-1$ can be expanded as
\br
\th^{(-1)} \eq \sum_{a=M+1}^{M+K} \th^{(-1)}_a E_{- \a_{a}}^{(0)}
+ \th^{(-1)}_{\psi} E_{\a_{M+1} +  \ldots +\a_{M+K}}^{(-1)} \nonu \\
&+& \sum_{l=1}^{M} \th^{(-1)}_l E_{-(\a_{l} +  \ldots +\a_{M+1})}^{(0)}
+ \sum_{l=1}^{M} {\bar \th}^{(-1)}_l E_{\a_{l} +  \ldots +\a_{M+K}}^{(-1)}
\lab{thm1}
\er
where we included all possible terms of grade $-1$ according to
\rf{grading}.

Therefore
\br
\sbr{E}{\th^{(-1)}} \eq \sum_{a=M+1}^{M+K} \th^{(-1)}_a \a_a \cdot H^{(0)}
+ \th^{(-1)}_{\psi} \( - (\a_{M+1} +  \ldots +\a_{M+K})\cdot H^{(0)} + c\)
\nonu \\
&+& \sum_{l=1}^{M} \th^{(-1)}_l \eps ( {\scriptstyle
  \a_{M+1} \,,\, -\a_{l} -  \ldots -\a_{M+1} } )\,
E_{-(\a_{l} +  \ldots +\a_{M})}^{(0)} \nonu \\
&+& \sum_{l=1}^{M} {\bar \th}^{(-1)}_l \eps ( {\scriptstyle
  -\a_{M+1} -  \ldots -\a_{M+K} \,,\, \a_{l} +  \ldots +\a_{M+K} } )\,
E_{\a_{l} +  \ldots +\a_{M}}^{(0)}
\lab{ethm1}
\er
Comparing the last expression with the field content of $A$ as given by
\rf{a20} we obtain relations for expansions parameters used in \rf{thm1}:
\br
&&\nu = - \th^{(-1)}_{\psi} \quad, \quad U_a = - \th^{(-1)}_{a} +
\th^{(-1)}_{\psi} \quad, \quad r_l = - \th^{(-1)}_l
\eps ( {\scriptstyle \a_{M+1} \, , \,-\a_{l} -  \ldots -\a_{M+1}  })
\nonu\\
&& q_l = - {\bar \th}^{(-1)}_l \eps (
{\scriptstyle -\a_{M+1} -  \ldots -\a_{M+K} \,, \,
\a_{l} +  \ldots +\a_{M+K}})
\lab{avsth}
\er
We now work with representation of $A$ as given in eq. \rf{abb}.
We split the grade zero element $B$ in a product $B= B_1 B_2 $
with $B_1$ containing the grade zero $sl (M)$ elements and
\be
B_2 \equiv \exp \sum_{a=1}^K \p_{M+a} \, \a_{M+a} \cdot H^{(0)} +
\rho \cdot {\hat c}
\lab{b2}
\ee
Accordingly, eq.\rf{abb} becomes
$A= - B_2^{-1} B_1^{-1} ( \pa_x B_1 ) B_2 - B_2^{-1} ( \pa_x B_2 )$
and $A$ can be rewritten as
\be
A = - \sum_{a=1}^K \pa_x \p_{M+a}\,  \a_{M+a} \cdot H^{(0)} -
\pa_x\rho \cdot {\hat c} + O (sl (M))
\lab{aslm}
\ee
where $O (sl (M))$ contains all possible terms belonging to the $sl (M)$
algebra.

Comparison with \rf{a20} yields
\be
U_{M+a} = - \pa_x \p_{M+a} \quad ;\quad \nu = - \pa_x\rho
\lab{umnu}
\ee

We define a family of the first order differential matrix
operators $\cL_N= \pa / \pa t_N - A_N$, $N=1, \ldots $.
The hierarchy is then formulated in terms of the zero-curvature equations
for the Lax operators
\be
\sbr{{\cL}_N}{{\cL}_M}=0
\lab{zer-curv}
\ee
expressing commutativity of the higher time flows.
The zero-curvature equations imply the pure gauge solutions for the
potentials $A_N$:
\be
\cL_N =  \Psi \partder{}{t_N} \Psi^{-1}
\lab{psidef}
\ee
The starting point of the dressing method \ct{FMSG} is the vacuum
solution $\nu = U=r=q=0$. The corresponding $L^{({\rm vac})} = D - E$ matrix
Lax
operator together with higher flows operators $\cL_N\, , N>1$ for the vacuum
solutions are expected to be recovered via \rf{psidef} from
$ \Psi$, which is  expressed entirely by the
Heisenberg algebra associated with the center of $\kere$.
Explicitly, for our model
\be
\Psi= \Psi^{({\rm vac})} = \exp \( \sumi{N=1} t_N b^{(N)}\) \lab{psivac}
\ee
with $b_{N}$ given in \rf{heisel}.

We define the  tau-function vectors as:
\be
\ket{\tau_{0} } \>  =\> \Psi^{({\rm vac})}\> h\> {\Psi^{({\rm vac})}}^{-1}\>
\ket{\l_0} \quad \; ; \quad \;
\ket{\tau_{M+a}} \>  =\> \Psi^{({\rm vac})}\> h\> {\Psi^{({\rm vac})}}^{-1}\>
\ket{\l_{M+a}}
\lab{tauvec}
\ee
They are associated with the constant group element $h$ and
the highest-weight vectors $\ket{\l_0},\ket{\l_{M+a}}$ such that
\be
\a_{M+a} \cdot H^{(0)} \ket{\l_0} =0
\quad \; ; \quad
\a_{M+b} \cdot H^{(0)} \ket{\l_{M+a}} = \d_{a,b} \ket{\l_{M+a}}
\quad ; \quad a,b = 1, \ldots ,K.
\lab{higwei}
\ee
Assuming that $h$ allows the ``Gauss'' decomposition of
$\Psi^{({\rm vac})}\> h\> {\Psi^{({\rm vac})}}^{-1}$
in positive, negative and zero grade elements we get for the tau-function
vectors from \rf{tauvec} an alternative expression:
\br
\ket{\tau_{0}} & =& \Th^{-1} \, B^{-1} \, \ket{\l_{0}} \; \; ; \; \;
\ket{\tau_{M+a}}  = \Th^{-1} \, B^{-1} \, \ket{\l_{M+a}}  \nonu \\
\Th^{-1} &=& \(\Psi^{({\rm vac})}\> h\> {\Psi^{({\rm vac})}}^{-1}\)_{-}
\;\; ; \;\;
B^{-1} = \(\Psi^{({\rm vac})} \> h\> {\Psi^{({\rm vac})}}^{-1}\)_{0}
\lab{alttau}
\er
for $a =  1, \ldots ,K$.
Like before, we make a splitting $B^{-1} = B_2^{-1} B_1^{-1}$ in \rf{alttau}
and notice that $B_1^{-1} \ket{\l_{M+a}} = \ket{\l_{M+a}}$
for $B_1$ being an exponential of $sl(M)$ generators.
Inserting $B_2$ from \rf{b2} we find
\be
\ket{\tau_{0}} \> = \Th^{-1} \,  \ket{\l_{0}}  \, e^{ \(- \rho \)}
\quad\; \; ; \; \; \quad
\ket{\tau_{M+a}} \> = \Th^{-1} \, B^{-1} \, \ket{\l_{M+a}} \, e^{\( - \rho -
\phi_{M+a} \)}
\lab{t02phi}
\ee
Denote
\be
\t^{(0)}_{0} \equiv \exp \(- \rho \) \qquad ; \qquad
\t^{(0)}_{M+a} \equiv \exp \(- \rho - \phi_{M+a} \)
\lab{tau00}
\ee
Accordingly, expanding $\Th^{-1}$ as below eq.\rf{bgamma} we find
\be
{ \ket{\t_{M+a}} \o \t^{(0)}_{M+a}} = \( 1 - \th^{(-1)} - \ldots \)
\ket{\l_{M+a}}
\lab{tmt0}
\ee
and similarly for $\ket{\t_{0}} / \t^{(0)}_{0}$.
We find by comparing with relation \rf{avsth} that
\br
r_l &=& \eps ( {\scriptstyle \a_{M+1} \,,\,-\a_{l} -  \ldots -\a_{M+1} }) \;
\langle \l_{M+1} \vert E_{\a_{l} +  \ldots +\a_{M+1}}^{(0)}
 \ket{\t_{M+1}} / \t^{(0)}_{M+1} \lab{rltau}\\
q_l &=&  \eps ({\scriptstyle  -\a_{M+1} -  \ldots -\a_{M+K}\, ,\,
  \a_{l} +  \ldots +\a_{M+K} }) \;
\langle \l_{0} \vert E_{-(\a_{l} +  \ldots +\a_{M+K})}^{(1)}
 \ket{\t_{0}} / \t^{(0)}_{0} \lab{qltau}\\
U_{M+a} &=& - \pa_x \ln \( { \t^{(0)}_{0}\, /\, \t^{(0)}_{M+a} }\) \qquad ;
\qquad \nu = - \pa_x \ln \(  \t^{(0)}_{0}\) \lab{uatau}
\er

The multi-soliton tau functions are defined in terms
of the constant group elements $h$ which are the product of
exponentials of eigenvectors of the Heisenberg subalgebra elements
\be
h = e^{F_1} \> e^{F_2} \> \cdots e^{F_n} \>, \qquad
[ b_N \> , \> F_k ] = \omega_N^{(k)} \> F_k \> , \quad k=1,2,
\ldots, n\>.
\lab{eigenb}
\ee
As seen from eq. \rf{eigenb} for such group elements the dependence of
the tau-vectors upon the times $t_N$ can be made quite explicit
\be
\ket{\tau_{a}} \>
=
 \prod_{k=1}^n \exp (e^{\sum_{N} \omega_{N}^{(k)} t_N} F_k )
 \> \ket{\l_a}
\lab{soltime}
\ee
The multi-soliton solutions are conveniently obtained in terms of
representations of the ``vertex operator'' type where the corresponding
eigenvectors are nilpotent.

\section{The $\bf sl(3)$ Example: Solitons of the Yaijma-Oikawa Hierarchy}
We apply the above method to the particular case of $sl(3)$ with $M=K=1$.
{}From eq.\rf{heisel} the surviving elements of the Heisenberg subalgebra are
in this case:
\br
b^{(2n+1)} &\equiv& b_{N=2n+1,a=1} = {\cal A}^1_{N=1+n\cdot 2}
=E^{(n)}_{\a_2} + E^{(n+1)}_{-\a_2}    \lab{b2n1}\\
b^{(2n)} &\equiv& b_{N=2n,0} = \sqrt{3} \l_1 \cdot H^{(n)}
- {{\hat c} \o 2 \sqrt{3}} \d_{n,0}   \lab{b2n}
\er
and they satisfy the usual Heisenberg subalgebra
$\sbr{b^{(k)}}{b^{(k^{\pr})}} = k \d_{k+k^{\pr}}$ for both even and odd $k$.

The structure of eigenvectors of Heisenberg subalgebra facilitates
construction of multisoliton solutions according to
\rf{eigenb} and \rf{soltime}.
In the current example we find that the eigenvectors and their corresponding
eigenvalues (in notation of \rf{eigenb}) are
\br
E_{{\ti \a}_1} &=& \sqrt{2} \sum_{n \in \IZ} \lb z^{-2n} E^{(n)}_{\a_1}
- z^{-2n-1} E^{(n)}_{\a_1+\a_2} \rb \quad ;\quad
\om^{(2n+1)}_{{\ti \a}_1} =  z^{2n+1} \quad ;\quad
\om^{(2n)}_{{\ti \a}_1} =  \sqrt{3} \, z^{2n}    \lab{eta1}\\
E_{{\ti \a}_2} &=&  \sum_{n \in \IZ} \llb z^{-2n-1} \(E^{(n)}_{\a_2}-
 E^{(n+1)}_{-\a_2} \)+ z^{-2n} \(\a_2 \cdot H^{(n)} - {{\hat c} \o 2}
\d_{n,0}\) \rrb   \lab{eta2}\\
&&\om^{(2n+1)}_{{\ti \a}_2} = - 2 z^{2n+1} \quad ;\quad
\om^{(2n)}_{{\ti \a}_2} = 0 \nonu \\
E_{{\ti \a}_1+{\ti \a}_2}\! &=& \!\sqrt{2} \sum_{n \in \IZ}
\lb z^{-2n} E^{(n)}_{\a_1} + z^{-2n-1}  E^{(n)}_{\a_1 +\a_2} \rb
\;\; ;\;\; \om^{(2n+1)}_{{\ti \a}_1+{\ti \a}_2} =  -z^{2n+1} \;\; ;\;\;
\om^{(2n)}_{{\ti \a}_1+{\ti \a}_2} = \sqrt{3}\, z^{2n}
\lab{eta12}
\er
We now realize the above eigenvectors by the nilpotent vertex operators.
The construction involves the Fubini-Veneziano operators defined in terms of
the Heisenberg elements as in eqs. \rf{qi}-\rf{qma}:
\be
Q_1 (z) \equiv i \sum_{n \in \IZ} { b^{(2n+1)} z^{-2n-1} \o 2n+1}
\quad ;\quad
Q_2 (z) \equiv q - ip \ln z+ i \sum_{n \ne 0} { b^{(2n)} z^{-2n} \o 2n}
\lab{fvqs}
\ee
where the zero mode momentum
$p = b^{(0)}=\sqrt{3} \l_1 \cdot H^{(0)} - {{\hat c} / 2 \sqrt{3}}$
satisfies $\sbr{q}{p}= i$.
The step operators from \rf{eta1}-\rf{eta12} are then realized, from the
algebra point of view, as vertex operators via:
\br
E_{{\ti \a}_1} &\leftrightarrow& E_{(1,\sqrt{3})} (z) =
\sqrt{2}\, z^{3/2} : \exp  \( i Q_1 (z) + i \sqrt{3}\, Q_2 (z) \):
\lab{vera1}\\
E_{{\ti \a}_2}  &\leftrightarrow& E_{(-2,0)} (z) =
- \h  : \exp  \( -2 i Q_1 (z) \): e^{i \pi p} \lab{vera2}\\
E_{{\ti \a}_1+{\ti \a}_2} &\leftrightarrow& E_{(-1,\sqrt{3})} (z) =
\sqrt{2} \, z^{3/2} : \exp  \( -i Q_1 (z) + i \sqrt{3}\, Q_2 (z) \):
\lab{vera3}
\er
and similarly for the negative root step operators, with change of sign of $i$
in exponentials.
A care has to be exercised in applying this correspondence within the setting
of the Fock space with the vacuum vector being $\ket{\l_2}$, since
$\langle \l_0 \v E_{{\ti \a}_2} \ket{\l_0} = - 1/2$
while
$\langle \l_2 \v E_{{\ti \a}_2} \ket{\l_2} =  1/2$ as seen from
expression \rf{eta2}.
Similar consideration applies for the products of
$E_{\a_i}$'s vertex operators like $E_{(-1,-\sqrt{3})} E_{(-1,\sqrt{3})}$
which produce $E_{(-2,0)}$.

Introduce the notation:
\be
V_{c_i, d_i} (z) \equiv z^{d_i^2/2}: \exp  \( i c_i \, Q_1 (z) + i d_i\,
Q_2 (z) \):
\lab{vcd}
\ee
It is not difficult to establish the following correlation function:
\br
&&\left\langle \l_{\s} \bigg\vert  \Psi^{({\rm vac})} V_{c_1, d_1} (z_1)
\cdots V_{c_n, d_n} (z_n)    {\Psi^{({\rm vac})} }^{-1}
\bigg\vert \l_{\s}  \right\rangle = \d_{\sum_{j=1}^n d_j,0} \;
e^{\sum_{j=1}^n  \G_{c_j, d_j} (z_j)  }
\nonu \\
&&\prod_{j=1}^n z_j^{(-1)^{({\s}+2)/2} {d_j \o 2 \sqrt{3}}
+{d_j^2\o 2} }
\prod_{1 \leq i < j \leq n} \( {z_i -z_j \o z_i+z_j} \)^{c_i c_j/2}
\llb \( z_i -z_j \) \( z_i+z_j\)\rrb^{d_i d_j/2}
\lab{corfun}
\er
for ${\s}=0,2$ and with
\be
\G_{c_j, d_j} (z_j) = \sumi{n=0} c_j \, t_{2n+1} z_j^{2n+1}+
\sumi{n=1} d_j \, t_{2n} z_j^{2n}
\lab{gamdef}
\ee
When substituting $V_{c_j, d_j} (z_j)$ by $E_{c_j, d_j} (z_j)$ one
encounters extra phases originating from the Klein factor in eq.\rf{vera2}
and from the character of the $\ket{\l_2}$ vacuum as discussed below
eq. \rf{vera3}.
The latter gives rise to the factor $\exp \( (i \pi /2) \sum_{j=1}^n c_j\)$
for the $\l_2$ correlation function as verified on several examples.

Recall that for the problem in hand the Lax matrix operator from \rf{lpsi}
with $A$ from \rf{a20} and $E$ from \rf{ele} specifies to
\be
L = D - \ttmat{0}{q}{0}
             {r}{U_2}{1}
             {0}{\l}{-U_2}
- \nu {\hat c}	
\lab{ttmat}
\ee
where $\l$ is the usual loop parameter.
In terms of the tau-vectors we have from \rf{rltau}-\rf{uatau}
the following $n$-soliton representation of the components of the Lax operator
\br
r &=& {1 \o  \t^{(0)}_{2}}  \langle \l_{2} \vert E_{\a_{1} + \a_{2}}^{(0)}
\ket{\t_{2}}
= {1 \o  \t^{(0)}_{2}}\,  \langle \l_{2} \vert E_{\a_{1} + \a_{2}}^{(0)}
\Psi^{({\rm vac})} \prod_{j=1}^n \( 1 + E_{c_j, d_j} (z_j) \)
{\Psi^{({\rm vac})} }^{-1} \vert \l_2  \rangle
\lab{rnsol} \\
q &=& {1 \o  \t^{(0)}_{0}}  \langle \l_{0} \vert E_{-\a_{1} - \a_{2}}^{(1)}
\ket{\t_{0}} 
= {1 \o  \t^{(0)}_{0}}  \langle \l_{0} \vert E_{-\a_{1} - \a_{2}}^{(1)}
\Psi^{({\rm vac})} \prod_{j=1}^n \( 1 + E_{c_j, d_j} (z_j) \)
{\Psi^{({\rm vac})} }^{-1} \vert \l_0  \rangle
\lab{qnsol} \\
U_2  &=& - \pa_x \ln \( { \t^{(0)}_{0} \, / \,  \t^{(0)}_{2} }\) \qquad ;
\qquad \nu = - \pa_x \ln \(  \t^{(0)}_{0}\) \lab{u2nsol}
\er
where
\be
\t^{(0)}_{\s} \,  = \, \langle \l_{\s} \vert \Psi^{({\rm vac})}
\prod_{j=1}^n \( 1 + E_{c_j, d_j} (z_j) \) {\Psi^{({\rm vac})} }^{-1}
\vert \l_{\s}  \rangle \quad ; \quad \s =0,2
\lab{tat}
\ee
Using association between the step operators \rf{eta1}-\rf{eta12}
and the vertex operators \rf{vera1}-\rf{vera3}
we can rewrite the step operators appearing in \rf{rnsol} and \rf{qnsol}
as
\be
E_{\a_{1} + \a_{2}}^{(0)} = - {1 \o 2 i \pi} \int d z_0 \,
V_{1, \sqrt{3}} (z_0)   \quad ; \quad
E_{-\a_{1} - \a_{2}}^{(1)}=  {1 \o 2 i \pi} \int d z_0 \,
V_{1, - \sqrt{3}} (z_0)
\lab{edz0}
\ee

We now calculate the zero-curvature equations \rf{zer-curv}
$ \sbr{D-E-A}{\pa_{t_n} -A_n} =0$ for the first two non-trivial cases of
$n=2,3$.
We expand $A_n = \sum_{i=0}^n A_n (i)$ where the index $i$ in the parenthesis
equals grading with respect to $\qs =  \l_{2} \cdot H^0 + 2 d$.
We choose $A_3 (3) = E_{\a_2}^{(1)} + E_{-\a_2}^{(2)}$ and
$A_2 (2) = \sqrt{3} \l_{1} \cdot H^{(1)}$ in order to ensure truncation of the
expansion.
This method yields for the first non-trivial case ($n=2$)
the evolution equations
\br
0 \eq \pa_t r +  \pa_x^2 r + r \pa_x U_2  - q r^2 - U_2^2 r
\;\; ; \; \;
0 = \pa_t U_2 +  \pa_x \( r q \)  \lab{t2a} \\
0 \eq \pa_t q  -  \pa_x^2 q + q \pa_x U_2  + r q^2 + U_2^2 q   \lab{t2q}
\er
where we defined for simplicity $t= \sqrt{3} t_2$.
The evolution equations for $t_3$ are
\br
0 &=& \pa_{t_3} U_2 - { 1 \o 4} \pa_x^3 U_2 + \h \pa_x U_2^3
+ { 3 \o 4} \pa_x \( r \pa_x q - q \pa_x r\)  \lab{t3a} \\
0 &=& \pa_{t_3} r -  \pa_x^3 r - {3 \o 2} \pa_x r \pa_x U_2 -
{3 \o 4} r \pa_x^2 U_2 +{3 \o 2}  r  U_2 \pa_x U_2 +
{3 \o 2}    U_2^2 \pa_x r \nonu \\
&+ &{3 \o 2}  q r^2  U_2
+{9 \o 4}  r q \pa_x r -{3 \o 4}  r^2 \pa_x q \lab{t3r} \\
0 &=& \pa_{t_3} q -  \pa_x^3 q + {3 \o 2} \pa_x q \pa_x U_2 +
{3 \o 4} q \pa_x^2 U_2 + {3 \o 2}  q  U_2 \pa_x U_2 +
{3 \o 2}    U_2^2 \pa_x q \nonu \\
&-& {3 \o 2}  q^2  r U_2
+{9 \o 4}  r q \pa_x q -{3 \o 4}  q^2 \pa_x r  \lab{t3q}
\er
These equations follow also from the conventional Sato equations
$ \pa_{t_n} \cL = \sbr{(\cL)^{(n/2)}_{+}}{\cL}$ applied to the scalar
cKP Lax operator $\cL = (\pa - U_2 )(\pa + U_2- q \pa^{-1} r )$.
We note here that the simple reduction of the matrix Lax
operator from \rf{ttmat} yields the scalar spectral problem
$\cL_1 \chi = \l \chi$ with the scalar Lax operator
$\cL_1 = (\pa + U_2 )(\pa - U_2- r \pa^{-1} q )$.
Both scalar Lax operators
are related by a conjugation and the Darboux-B\"{a}cklund transformation:
$\cL_1 = (\pa + U_2 ) \cL^{\ast} (\pa + U_2 )^{-1}$.

We now present few examples of the soliton solutions \rf{rnsol}-\rf{tat}
satisfying the above evolution equations.

\nit
1) \underbar{One-soliton solution, $n=1$.} With $h = ( 1 + E_{-2,0} (z_1))$
we recover the standard m-KdV one-soliton configuration with $r=q=0$ and
\be
\t^{(0)}_0 =  1 - \h \, e^{(\, - 2\,{x}\,{ z_1} - 2\,{ t_3}\,
{ z_1}^{3}\,)} \quad ;\quad
\t^{(0)}_2 =  1 + \h \, e^{(\, - 2\,{x}\,{ z_1} - 2\,{ t_3}\,
{ z_1}^{3}\,)}
\lab{yes}
\ee

\nit
2) \underbar{Two-soliton solutions, $n=2$.}
For $h = ( 1 + E_{-2,0} (z_1))( 1 + E_{1,\sqrt{3}} (z_2))$ we find
$\t^{(0)}_0$ and $\t^{(0)}_2$ as in \rf{yes} but now with $q \ne 0$ and
equal to
\be
q = - \sqrt{2}\,\,{z_2}\,e^{(\,{ t_3}\,{z_2}^{3} + {t}\,{z_2}^{2} + {x}\,
{z_2}\,)}\, \left(  \,1
 + \h\,{e}^{(\, - 2\,{x}\,{z_1} - 2\,{t_3}\,{ z_1}^{3}\,)}\,\({z_1+ z_2\o z_1
 -  z_2} \)  \, \!  \right) / \t^{(0)}_0
\lab{qyes}
\ee
while $r=0$.
Similarly, for $h = ( 1 + E_{-2,0} (z_1))( 1 + E_{1,-\sqrt{3}} (z_2))$ we find
$r \ne 0$ but $q =0$.

For $h = ( 1 + E_{1,\sqrt{3}} (z_1))( 1 + E_{1,-\sqrt{3}} (z_2))$ we find
that both $q \ne 0$ and $r \ne 0$:
\be
\t^{(0)}_{\s} = 1 + (-1)^{(\s/2)} \,2
\,{\displaystyle \frac {{ z_1}^{1+\s/2}\,{ z_2}^{2-\s/2}\,{\rm e}^{(\,{x}\,
{ z_1} + {t}\,{ z_1}^{2} +t_3 z_1^3+ {x}\,{ z_2} - {t}\,{ z_2}^{2}+t_3 z_2^3
\,)}}{(\,{ z_1} - { z_2}\,)\,(\,{ z_1} + { z_2}\,)^{2}}}
\quad ; \quad \s =0,2
\lab{t02sat}
\ee
\be
r = \frac{\sqrt{2}\,{ z_2}\,{\rm e}^{(\,
 - {t}\,{ z_2}^{2} + {x}\,{ z_2}+t_3 z_2^3\,)}}{\t^{(0)}_2}
\qquad ; \qquad
q=  \frac{\sqrt{2}\,{ z_1}\,{\rm e}^{(\,{t
}\,{ z_1}^{2} + {x}\,{ z_1}+t_3 z_1^3\,)}}{\t^{(0)}_0 }
\lab{rqsat}
\ee

\nit
3) \underbar{Three-soliton solutions, $n=3$.}
As an example we take here
$h = ( 1 + E_{-2,0} (z_1))( 1 + E_{1,\sqrt{3}} (z_2))
( 1 + E_{1,-\sqrt{3}} (z_3))$.
We find
\br
\t^{(0)}_{\s}
&= & 1 +(-1)^{(\s/2)}\, \h \,{e}^{(\, - 2\,{x}\,{ z_1} - 2\,{ t_3}\,{ z_1}^{3}\,)}
\nonu \\
& +& (-1)^{(\s/2)}\, 2\,{     {z_3}^{2-\s/2}\,{ z_2}^{1+\s/2}\,
 \o  {(\,{ z_2} - { z_3}\,)\,(\,{ z_2} + { z_3}\,
)^{2}} }
{e}^{(\,{ t_3}\,{ z_2}^{3} + {t}\,{ z_2}^{2} + {x}\,
{ z_2} + { t_3}\,{ z_3}^{3} - {t}\,{ z_3}^{2} + {x}\,
{ z_3}\,)}   \lab{t02tues} \\
&+ &
{   {z_3}^{2-\s/2}\,{ z_2}^{1+\s/2}\,(\,{ z_1} + { z_2}\,)\,(\,
{ z_1} + { z_3}\,)
\o
\,(\,{ z_2} - { z_3}\,)\,(\,{ z_2} + { z_3}\,)^{2}\,(\,
{ z_1} - { z_2}\,)
 (\,{ z_1} - { z_3}\,)\, }
 {e}^{(\, - 2\,{x}\,{ z_1} - 2\,{ t_3}\,{ z_1}^{3}
 + {t}\,{ z_2}^{2} + {x}\,{ z_2} + { t_3}\,{ z_2}^{3} - {
t}\,{ z_3}^{2} + {x}\,{ z_3} + { t_3}\,{ z_3}^{3}\,)}
\nonu \\
r &= &
\sqrt{2}\,{ z_3}\,{\rm e}^{(\, - {t}\,{ z_3}^{2
} + {x}\,{ z_3} + { t_3}\,{ z_3}^{3}\,)}\, \left( \! \,1 +
{\displaystyle \frac {1}{2}}\,{\displaystyle \frac {{\rm e}^{(\,
 - 2\,{x}\,{ z_1} - 2\,{ t_3}\,{ z_1}^{3}\,)}\,(\,{ z_1}
 + { z_3}\,)}{{ z_1} - { z_3}}}\, \!  \right) /\t^{(0)}_2
\lab{rtues} \\
q &= &
\sqrt{2}\,{z_2}\,{\rm e}^{(\,{t}\,{z_2}^{2}
 + {x}\,{z_2} + {t_3}\,{z_2}^{3}\,)}\, \left( \! \,1 -
{\displaystyle \frac {1}{2}}\,{\displaystyle \frac {{\rm e}^{(\,
 - 2\,{x}\,{z_1} - 2\,{t_3}\,{z_1}^{3}\,)}\,(\,{z_1}
 + {z_2}\,)}{{z_1} - {z_2}}}\, \!  \right) /\t^{(0)}_0
\lab{qtues}
\er

In the above examples $U_2$ and $\nu$ can be obtained from \rf{u2nsol}.
We note that we only kept the explicit time dependence on times $t_n$
with $n \leq 3$, which
was enough to verify the evolution equations \rf{t2a}-\rf{t3q}.

The novel feature of the above soliton solutions
is that they mix exponentials $\exp \( \sumi{n=1} t_n z_j^n\)$
which represent a typical time dependance for the KP solutions
with pure KdV-like time dependance of the type
$\exp \( \sumi{n=0} t_{2n+1} z_j^{2n+1}\)$ involving only odd times.

The exception is provided by the pure KP type of solution in eq.\rf{t02sat}-
\rf{rqsat}, which can also be obtained as a Wronskian arising from the Darboux
-B\"{a}cklund transformations.

\section{Conclusions}
We presented here new soliton solutions for the cKP hierarchy.
They exhibit a nontrivial mixing of KP times $t_n$ with all indices $n$'s and
KdV-like times $t_{2n-1}$ with only odd indices.
{}From the point of view of intermediate gradation, such a mixture is indeed
very natural. In fact, we are able to obtain pure KdV solutions, pure KP
solutions (meaning all times) and the arbitrary mixtures thereof,
just by varying the constant group elements $h$ of the dressing orbit.

We were not able to obtain these solutions from the conventional
Darboux-B\"{a}cklund method involving Wronskian representation.
Recall, that such Wronskian is given in terms of the eigenfunctions $f_i$
satisfying $\pa_n f_i = \pa_x^n f_i$ for all $n$.
This type of time dependence fits only a subclass of our soliton
solutions, namely those which are of the pure KP type
(meaning that all $t_n$'s with all
$n$'s are present without the nontrivial mixture with KdV times).
Based on studying many examples of these solutions, we believe that
the Wronskian method works equally well in these cases.

\lskip
{\bf Acknowledgements }
HA was supported in part by the U.S. Department of Energy Grant No.
DE-FG02-84ER40173.
A.H.Z. would like to thanks UIC and FAPESP
for the hospitality and financial support. L.A.F., J.F.G. and A.H.Z.
acknowledge partial financial support from CNPq.
\lskip

\end{document}